\documentclass[onecolumn,showpacs,showkeys,preprintnumbers]{revtex4}
\usepackage{amssymb}

\usepackage{amsmath}
\usepackage{graphicx}

\begin{document}

\title{The Integrated Sachs-Wolfe Effect in Time Varying Vacuum Model}
\author{Y.T. Wang, Y.X. Gui, L.X. Xu}
\email{lxxu@dlut.edu.cn}
\author{J.B. Lu} \affiliation{School of Physics and Optoelectronic Technology,\\
Dalian University of Technology, Dalian, Liaoning 116024, P. R.
China}
\begin{abstract}
The integrated Sachs-Wolfe (ISW) effect is an important implication
for dark energy. In this paper, we have calculated the power
spectrum of the ISW effect in the time varying vacuum cosmological
model, where the model parameter $\beta=4.407$ is obtained by the
observational constraint of the growth rate. It's found that the
source of the ISW effect is not only affected by the different
evolutions of the Hubble function $H(a)$ and the dimensionless
matter density $\Omega_m(a)$, but also by the different growth
function $D_+(a)$, all of which are changed due to the presence of
matter production term in the time varying vacuum model. However,
the difference of the ISW effect in $\Lambda(t)\textmd{CDM}$ model
and $\Lambda \textmd{CDM}$ model is lessened to a certain extent due
to the integration from the time of last scattering to the present.
It's implied that the observations of the galaxies with high
redshift are required to distinguish the two models.
\end{abstract}

 \maketitle

\address{Department of Physics, Dalian University of Technology,}

\section{Introduction}

Since 1998, the type Ia supernova (SNe Ia) observations
\cite{ref:Riess98,ref:Perlmuter99} have showed that the expansion of
our universe is speeding up rather than slowing down. During these
years from that time, many additional observational results,
including current Cosmic Microwave Background (CMB) anisotropy
measurement from Wilkinson Microwave Anisotropy Probe (WMAP)
\cite{ref:Spergel03,ref:Spergel06}, and the data of the Large Scale
Structure (LSS) \cite{ref:Tegmark1,ref:Tegmark2}, also strongly
support this suggestion. In order to understand the mechanism of the
accelerating expansion of the universe, the common recognition has
been received that there exists an exotic energy component with
negative pressure, called dark energy (DE), whose density accounts
for two-thirds of the total energy density in the universe according
to the current observations. It has been the most important thing to
make an effort to probe into what the nature of DE is in the modern
cosmology. The simplest and naturalest candidate is the cosmological
constant ${\Lambda}$ (vacuum energy) \cite{ref:LCDM1,ref:LCDM2},
with the equation of state (EOS) $w=\frac{P}{\rho}=-1$. However,
it's pity that it confronts with two difficulties: the fine-tuning
problem and the cosmic coincidence problem. An alternative offer is
the dynamical DE to alleviate the difficulties above, such as
quintessence \cite{ref:quintessence}, phantom \cite{ref:phantom},
quintom \cite{ref:quintom}, generalized Chaplygin gas (GCG) model
\cite{ref:GCG}, Holographic DE \cite{ref:holo}, and so on.

At present, although in the presence of numerous dark energy models,
the concordance model still fits best well with the current combined
observations \cite{ref:WMAP5}. So it's necessary to review the
cosmic constant problems: The fine-tuning problem is the vacuum
energy density predicted by quantum field theory (QFC) is
$10^{74}GeV^4$, which is about $10^{121}$ orders of magnitude larger
than the observed value $10^{-47}GeV^4$, so there must be so-called
"fine-tuning" while looking for an balance between two numbers which
exits so much discrepancy. The cosmic coincidence problem says:
Since the energy densities of vacuum energy and dark matter turn on
different evolutions during the expansion history of the universe,
why are they nearly equal today and why it happens just now? To get
this coincidence, it appears that their ratio must be set to a
specific, infinitesimal value in the very early universe. Therefore,
in nature the second difficulty is also "fine-tuning" problem. In
the final analysis, both of the two difficulties are related to the
energy density of vacuum energy. On the basis of this consideration,
a phenomenological and natural attempt at alleviating the above
problems is allowing vacuum energy $\Lambda$ to vary
\cite{ref:LCDM2, ref:LtCDM}. In $\Lambda(t)\textmd{CDM}$ model, we
keep unvaried for the EOS of the vacuum energy, equaling to $-1$,
and have an time-evolving energy density. Although the weak point in
this ideology is the unknown functional form of the $\Lambda(t)$
parameter, we can deal with the $\Lambda(t)$ parameter by different
phenomenologically parameterized form, referring to
\cite{ref:LCDM2,ref:LtCDMparam1}. Meanwhile, the underly explanation
for $\Lambda(t)$ parameter has also been explored in the framework
of QFT using the renormalization group \cite{ref:RG}.

The ISW effect \cite{ref:ISW1} arises from a time-dependent
gravitational potential at late time and leads to CMB temperature
anisotropies. From the physical picture point of view, the frequency
of the photon released by CMB has been redshifted or blueshifted,
which is caused by the decaying gravitational potential, while
propagating along the line of light-like from the last scattering
surface to now. The ISW effect is a significant implication for dark
energy. Since the gravitational potential was an constant during
matter-dominated. At the late time, the gravitational potential is
time-dependent when dark energy begin to dominate in the universe.
So, the detection of ISW effect will provide great support and more
information for dark energy. Now, the ISW effect is detected by the
cross-correlation between CMB and LSS observations
\cite{ref:LSS1,ref:densitycontrast,ref:LSS3,ref:LSS4,ref:LSS5}. In
this way, we can separate CMB temperature anisotropies caused by ISW
effect from the primary temperature anisotropies.

In this paper, we consider the time varying vacuum model, where a
parameterized form of $\Lambda(t)$ is used. The densities of dark
matter and vacuum energy don't conserve independently owing to the
time varying vacuum term, which plays the role of the interaction
between dark matter and dark energy. The matter production term
leads to the different evolutions of the Hubble parameter $H(a)$ and
the dimensionless matter density $\Omega_m(a)$, and the different
structure formation, comparing with the concordance model.
Meanwhile, there being a relation between these three quantities and
the source term of the ISW effect, we explore the impact from the
change of these three quantities on ISW effect in the
$\Lambda(t)\textmd{CDM}$ model in theory.

The paper is organized as follows. In next section, we briefly
review of the cosmological evolution and structure growth in time
varying vacuum model. In section III, we focus on the theory and
detection of ISW effect and its power spectrum. The last section is
the conclusion.

\section{Review of the cosmological evolution and structure growth in Time Varying Vacuum Model }

Considering a spatially flat Friedmann-Robertson-Walker(FRW)
cosmological model, which consists of two components: the dark
matter and the dynamic DE. With the metric
$ds^2=-dt^2+a^2(t)[dr^2+r^2(d{\theta}^2+\sin^2{\theta}d{\phi}^2)]$,
the Friedmann equation and energy-momentum conservation equation can
be written as
\begin{eqnarray}
&&H^2=\frac{1}{3M_{pl}^2}({\rho}_m+{\rho}_{\Lambda}),\\
&&\frac{d(\rho_m+\rho_{\Lambda})}{dt}=-3H(\rho_m+P_m+\rho_{\Lambda}+P_{\Lambda}),
\end{eqnarray}
where, $H=\frac{\dot{a}}{a}$ is the Hubble parameter and
$M_{pl}\equiv (8{\pi}G)^{-\frac{1}{2}}$ is the reduced Planck mass.
${\rho}$ and $P$ are density and pressure of a general piece of
matter, and their subscripts $m$ and $\Lambda$ respectively
correspond to dark matter and dark energy.

In the present paper, we consider the time vacuum energy as dynamic
dark energy with energy density $\rho_{\Lambda}=\Lambda(t)M_{pl}^2$.
In order to simplify the formalism, we work in the framework of
geometrical units with $8{\pi}G=c\equiv1$. Thus, we can obtain the
vacuum energy density $\rho_\Lambda=\Lambda(t)$ and the
corresponding equation of state
$P_\Lambda=-\rho_\Lambda=-\Lambda(t)$. Then the equation (2) reads
\begin{eqnarray}
&&\frac{d\rho_m}{dt}+3H\rho_m=-\dot{\Lambda}(t),
\end{eqnarray}
which shows that the decaying vacuum density $\Lambda(t)$ plays the
role of a source of matter production, and the densities of dark
matter and vacuum energy don't conserve independently. Now, we
consider a parameterized form using a power series of $\Lambda(t)$
\cite{ref:LtCDMparam1, ref:LtCDMparam2}, this is :
\begin{eqnarray}
&&\Lambda(t)=n_1H+n_2H^2.\label{4Ltpara}
\end{eqnarray}
Combining the equations (1), (3) and (4), we can get the following
Hubble function:
\begin{eqnarray}
&&H(t)=\frac{n_1}{\beta}\frac{e^\frac{n_1t}{2}}{e^\frac{n_1t}{2}-1},
\end{eqnarray}
where we have defined $\beta=3-n_2$. Then with the definition of
Hubble function $H(t)=\frac{\dot{a}}{a}$, we proceed with the
integration, deriving the evolution of the scale factor of the
universe $a(t)$:
\begin{eqnarray}
&&a(t)=a_1(e^\frac{n_1t}{2}-1)^\frac{2}{\beta},
\end{eqnarray}
where $a_1$ is the constant of integration. Considering the two
equations above, we obtain after some algebra that the Hubble
function evolves with the scale factor of the universe as follows:
\begin{eqnarray}
&&H(a)=\frac{n_1}{\beta}\left[1+\left(\frac{a(t)}{a_1}\right)^{-\frac{\beta}{2}}\right],
\end{eqnarray}
where, the parameter $n_1$ and the integration constant $a_1$ can be
reexpressed by taking the Hubble constant $H_0$ and the current
value of dimensionless dark matter density
$\Omega_m(a=1)=\Omega_{m0}$ into account with the help of equations
(1), (4) and (7):
\begin{eqnarray}
&&n_1=\frac{\beta
H_0}{1+a_1^\frac{\beta}{2}},~~~~~~~a_1=\left(\frac{3\Omega_{m0}}{\beta-3\Omega_{m0}}\right)^\frac{\beta}{2}.
\end{eqnarray}

 Then making use of equation (8), we evaluate the dimensionless dark
matter density $\Omega_m(a)$ and the dimensionless quantity $E(a)$:
\begin{eqnarray}
&&\Omega_m(a)=\frac{\Omega_{m0} a^\frac{-\beta}{2}}{E(a)},\\
&&E(a)=1-\frac{3\Omega_{m0}}{\beta}+\frac{3\Omega_{m0}}{\beta}a^\frac{-\beta}{2}.
\end{eqnarray}
From equations (9) and (10), it's obvious that the Hubble parameter
$H(a)$ and the matter density $\Omega_m(a)$ scale differently,
respectively comparing with their evolutions in the concordance
model, due to the decaying vacuum density. In addition, we can get
the comoving distance $\chi(a)$ from the solution of the Hubble
function $H(a)$ according to its definition of the integration :
\begin{eqnarray}
&&\chi(a)=\int_a^1\frac{da}{a^2H(a)}.
\end{eqnarray}

In the following, we consider the structure growth in the framework
of time varying vacuum model. The corresponding Newtonian equation
governing the time evolution of dark matter perturbation could be
generalized by taking a source of matter production into account
\cite{ref:growth}. It can be written as:
\begin{eqnarray}
&&\ddot{D}_++(2H+\frac{\epsilon}{\rho_m})\dot{D}_+-\left[\frac{\rho_m}{2}-
2H\frac{\epsilon}{\rho_m}-\dot{\left(\frac{\epsilon}{\rho_m}\right)}\right]D_+=0,
\end{eqnarray}
where we have defined a growth function
$D_+(a)=\frac{\delta(x,a)}{\delta(x,a=1)}$, an equivalent and
convenient quantity describing the growth of dark matter density
inhomogeneities. $\epsilon$ is the source of matter production, and
here it is given by $\epsilon=-\dot{\Lambda}(t)$. On account of the
presence of $\epsilon$, it's seen that equation (12) is different
from the well-known linear matter perturbation equation.

There are two solutions to the differential equation (12), where one
is a growing mode and the other is a decaying mode. Since almost all
current models of the universe have a non-increasing Hubble rate, we
are interested in the growing one, which is given by
\cite{ref:LtCDMparam1, ref:LtCDMparam2}:
\begin{eqnarray}
&&D_+(a)=\frac{9\Omega_{m0}^2}{2\beta}\left(\frac{\beta-3\Omega_{m0}}{3\Omega_{m0}}\right)
^\frac{2\beta-4}{\beta}a^\frac{\beta-6}{2}\int_0^a\frac{dx}{x^\frac{\beta}{2}
E^2(x)}.
\end{eqnarray}

Next, we constrain the $\Lambda(t) CDM$ model parameter $\beta$
according to the growth rate of clustering:
\begin{eqnarray}
&&f(a)=\frac{a}{D_+(a)}\frac{d D_+(a)}{d a}.
\end{eqnarray}
\begin{table}
\begin{center}
\begin{tabular}{cc   cc   cc}
\hline\hline redshift z && $f_{obs}$  &  & Reference &
\\ \hline
$0.15$        && $0.51\pm0.11$ &     & \cite{ref:fobs1} &  \\
$0.35$        && $0.70\pm0.18$ &     & \cite{ref:fobs2} &  \\
$0.55$        && $0.75\pm0.18$ &     & \cite{ref:fobs3} &  \\
$0.77$        && $0.91\pm0.36$ &     & \cite{ref:fobs4} &  \\
$1.4$         && $0.90\pm0.24$ &     & \cite{ref:fobs5} &  \\
$3.0$         && $1.46\pm0.29$ &     & \cite{ref:fobs6} & \\
\hline\hline
\end{tabular}
\caption{The currently available observational data for the growth
rate of clustering and its measurement
uncertainty.}\label{tab:fobsdata}
\end{center}
\end{table}

\begin{figure}[!htbp]
\includegraphics[width=8cm]{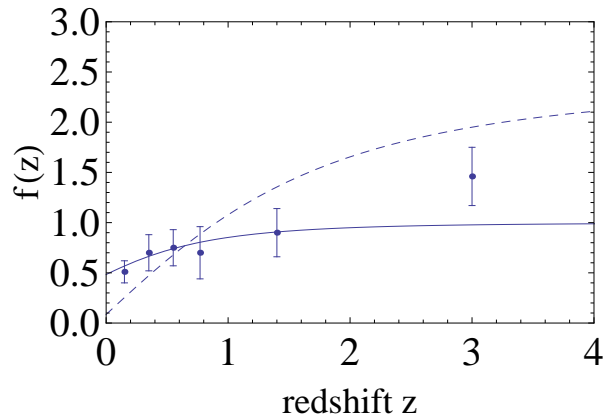}\\
  \caption{The observed values of the growth rate $f_{obs}$ with $1\sigma$ uncertainty are plotted by the dots with error bars.
  The theoretical growth rate $f_{th}$ evolve with the redshift $z$ for $\Lambda \textmd{CDM}$ model (solid line) and for the $\Lambda(t)\textmd{CDM}$
  (dashed line).
 }\label{fig:growthrate}
\end{figure}

Using the currently available growth rate data, as shown in Table
\ref{tab:fobsdata}, we can determine the best fit value of the model
parameter $\beta$ by minimizing
\begin{eqnarray}
&&\chi^2(\beta)=\sum_i\left[\frac{f_{obs}(z_i)-f_{th}(z_i;\beta)}{\sigma_{f_{obs}}(z_i)}\right]^2,
\end{eqnarray}
where $f_{obs}$ is the observed growth rate, and $\sigma_{f_{obs}}$
is its corresponding $1\sigma$ measurement uncertainty. $f_{th}$ is
the theoretical value and can be obtained from equations (13) and
(14). Here, we have taken a prior on $\Omega_{m0}=0.27$
\cite{ref:WMAP5}. The best fit value of model parameter
$\beta=4.407$ is obtained with the minimization $\chi^2$
$(\chi^2_{min}/dof=2.83)$. In FIG.\ref{fig:growthrate}, we plot the
observed values of the growth rate and its theoretical evolutions in
$\Lambda \textmd{CDM}$ model and $\Lambda(t)\textmd{CDM}$ model with
the prior value of $\Omega_{m0}=0.27$ and the best fit value of
model parameter $\beta=4.407$. From FIG. \ref{fig:growthrate}, it's
found that the theoretical evolution of the growth rate in the
$\Lambda \textmd{CDM}$ model has a minor difference from the
observational data at low redshifts.

\begin{figure}[!htbp]
  \includegraphics[width=8cm]{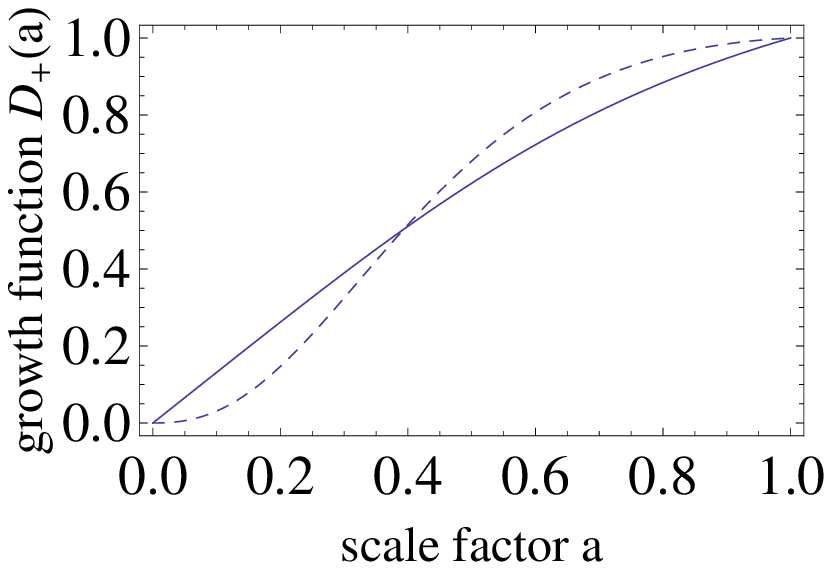}
  \includegraphics[width=8cm]{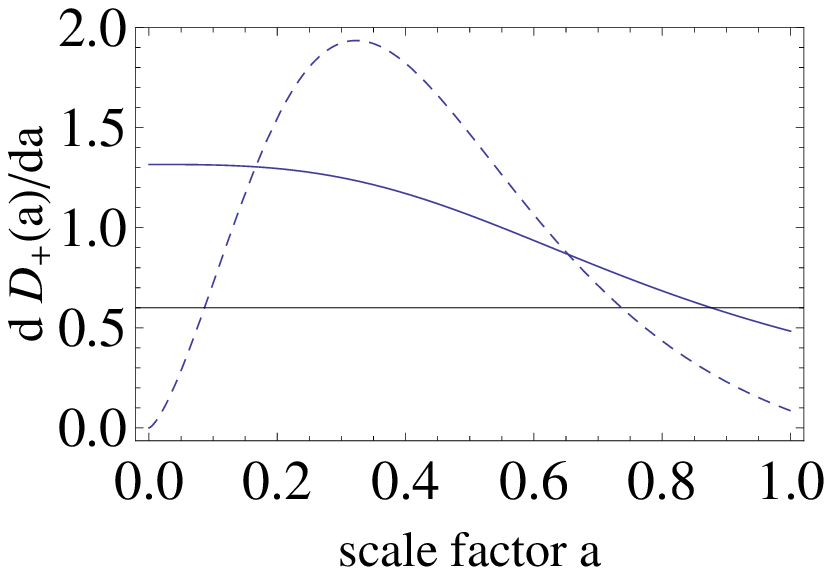}\\
  \caption{$D_+(a)$ vs. $a$ for $\Lambda \textmd{CDM}$ model (solid line) with $\Omega_{m0}=0.27$ and
  $\Lambda(t)\textmd{CDM}$ model (dashed line) with $\Omega_{m0}=0.27$ and $\beta=4.407$ on the left.
  The right panel shows $dD_+(a)/da$ scales with the scale factor $a$: the solid line is for $\Lambda \textmd{CDM}$ model;
  the dashed line is for $\Lambda(t)\textmd{CDM}$ model. The values of the cosmological
  parameter $\Omega_{m0}=0.27$ and the Hubble constant $H_0=70.5km/s/Mpc$
  are taken by \cite{ref:WMAP5}.
  }\label{fig:D}
\end{figure}

Then, the evolutions of growth function $D_+(a)$ and its change rate
$dD_+(a)/da$ as a function of the scale factor have been plotted in
FIG.\ref{fig:D}. From FIG.\ref{fig:D}, we can see the growth
function in $\Lambda(t)\textmd{CDM}$ model is different from that in
$\Lambda \textmd{CDM}$ model for the presence of time varying vacuum
energy. And it's found that $D_+(a)$ is a little smaller than
$D_\Lambda$ at $a<0.4$ in the left panel, then there is a trend that
the growth function in $\Lambda(t)CDM$ model keeps larger than
$D_\Lambda$ until now. In the right panel of FIG. \ref{fig:D}, it's
found that when $a<0.16$ and $a>0.65$, the rate of change in the
growth function $D_+(a)$ in $\Lambda(t)\textmd{CD}M$ model is slower
than that in $\Lambda \textmd{CDM}$ model, but quicker than the
change rate of $D_\Lambda$ in $\Lambda \textmd{CDM}$ model with the
scale factor ranges $0.16<a<0.65$. And we can find that the
increasing rate of the growth function $D_{\Lambda}(a)$ appears to
be more and more small along all the scale factor. The right panel
shows that there exists an upward trend in the increasing rate of
the growth function $D_+(a)$ of $\Lambda(t)\textmd{CDM}$ model till
$a=0.3$, then it gradually starts to decrease. In
FIG.\ref{fig:growthrate} and FIG. \ref{fig:D}, it's indicated that
there is much discrepancy between the structure growth history in
the two models.

\section{Theory of ISW effect and angular power spectrum}
In this section, we give a brief review on ISW effect and its
detection, and calculate its angular power spectrum. The ISW effect
arises from a time-dependent gravitational potential at late time
when dark energy dominates in the universe. The CMB temperature
perturbation caused by ISW effect \cite{ref:ISW2} is expressed by
\begin{eqnarray}
&&\frac{\Delta T}{T_{CMB}}\equiv2\int
d\tau\frac{\partial\Phi}{\partial\tau}=-2\int_0^{\chi_H} d\chi
a^2H(a)\frac{\partial\Phi}{\partial a},
\end{eqnarray}
where $\tau$ and $\chi$ are respectively the conformal time and
comoving distance, which are related by $d\chi=-d\tau$. $\Phi$
denotes the gravitational potential, which is related to dark matter
perturbation $\delta$ according to Poisson equation on sub-horizon
scales:
\begin{eqnarray}
\nabla^2\Phi(x)=\frac{a^2\rho_m(a)\delta(x)}{2},
\end{eqnarray}
Carrying out Laplace transform for the two sides of equation (17)
and taking the dimensionless dark matter density $\Omega_m(a)$ into
consideration, we obtain:
\begin{eqnarray}
&&\Phi(k,a)=-\frac{3}{2}a^2H^2(a)\Omega_m(a)\frac{\delta(k,a)}{k^2}.
\end{eqnarray}
Substituting the above equation into equation (16), we are able to
write down the CMB temperature perturbation coming from ISW effect:
\begin{eqnarray}
&&\frac{\Delta T}{T_{CMB}}=3H_0^2\int_0^{\chi_H}d\chi
a^2H(a)\frac{d\zeta(a)}{da}\frac{\delta(k,a=1)}{k^2},
\end{eqnarray}
where the redefined function $\zeta(a)=a^2E^2(a)\Omega_m(a)D_+(a)$
is used.

From equation (19), we can see that the source term of ISW effect
depends on the following three quantities: the Hubble function
$H(a)$, the dimensionless matter density $\Omega_m(a)$ and growth
function $D_+(a)$, all of which are affected by the dark matter
production term as we have mentioned in section II.

From the observation point of view, ISW effect leads to these "new"
temperature anisotropies, smaller than those produced at the last
scatting surface. So there are many difficulties in directly
detecting this effect apart from primary CMB temperature
anisotropies. Now, this embarrassment is solved by searching for the
cross-correlation between ISW temperature and LSS observations,
which include the galaxies sample \cite{ref:SDSSmain, ref:galaxies},
quasars sample \cite{ref:SDSSquasars}, and Luminous Red Galaxies
(LRG)\cite{ref:SDSSLRG} from SDSS, the radio galaxy catalogue from
the NRAO VLA Sky Survey (NVSS) \cite{ref:NVSS}, the infrared
catalogue from two Micron All-Sky Survey (2MASS) \cite{ref:2MASS},
and the X-ray catalogue from the High Energy Astrophysical
Observatory (HEAO) \cite{ref:HEAO}. Here, we use the samples from
SDSS and NVSS, which span different redshift ranges. The more
detailed presentation on the characteristics of all the samples are
given in Ref. \cite{ref:densitycontrast,ref:LSS3}. The observed
galaxies density contrast will be
\cite{ref:densitycontrast,ref:LSS3}:
\begin{eqnarray}
&&\delta_g=\int b(z)\frac{d N}{dz}\delta(z)dz=\int f(z) \delta(z)dz,
\end{eqnarray}
where $\frac{d N}{dz}$ is the selection function of the survey,
$b(z)$ is galaxy bias relating the visible matter distribution
$\delta_g$ to the underlying dark matter $\delta(z)$, which is
either a constant or time-evolving with a parameterized form
\cite{ref:bias}. Here we adopt that it is scale independent for
simplicity.

\begin{figure}[!htbp]
\includegraphics[width=12cm]{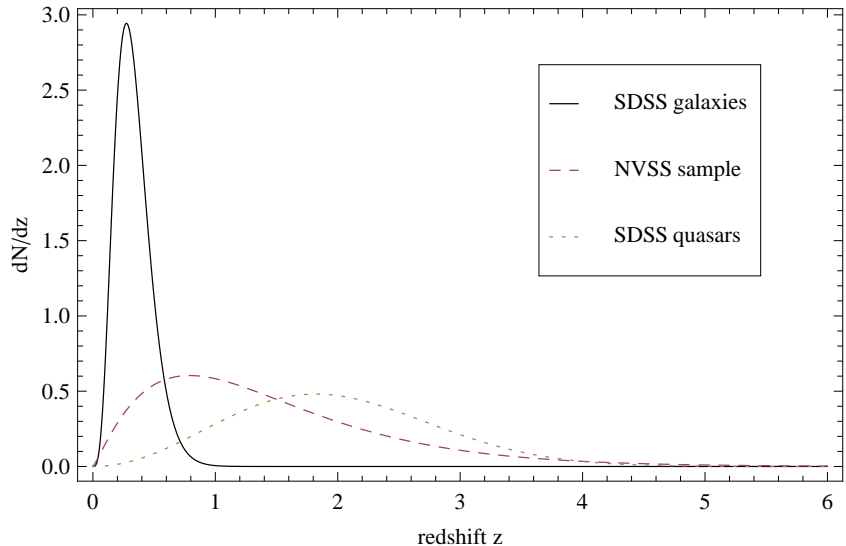}\\
  \caption{the theoretical redshift distributions d N/d z of three
  samples with different redshift ranges.
 }\label{fig:reddis}
\end{figure}

We use the approximate function of the redshift distribution of the
observed samples released by
\cite{ref:densitycontrast,ref:LSS3,ref:ISW2, ref:RedD, ref:sample}:
\begin{eqnarray}
&&f_{SDSS}(z)=b(z)\frac{d
N}{dz}=b_g\frac{\gamma}{\Gamma((m+1)/\gamma)}\frac{z^m}{z_0^{m+1}}\exp\left[-\left(\frac{z
}{z_0}\right)^\gamma\right], \\
&&f_{NVSS}(z)=b(z)\frac{d
N}{dz}=b_{eff}\frac{\alpha^{\alpha+1}}{z_{\ast}^{\alpha+1}\Gamma(\alpha)}z^{\alpha}e^{-\alpha
z/z_{\ast}},
\end{eqnarray}
where the parameters can be obtained according to the observation of
the galaxy number density after we have selected a certain form of
redshift distribution. For the SDSS galaxies sample with the
redshift range $0.1<z<0.9$, the best fittings of parameters in
equation (19) are given by $z_0= 0.113$, $\gamma=1.197$ and
$m=3.457$ with a median redshift of $z_{med}=0.32$ and a constant
bias $b_g=1$ in Ref. \cite{ref:densitycontrast}. For the NVSS sample
with a little wider redshift range, the best values of parameters in
equation (20) are given by the effective bias $b_{eff}=1.98$,
$z_{\ast}=0.79$ and $\alpha=1.18$ in Ref. \cite{ref:LSS3}. For the
SDSS quasars sample with the redshift between 0.065 and 6.075, it's
found that $z_0=1.9$, $\gamma=2.2$ and $m=2$ and its median redshift
$z_{med}\sim1.8$\cite{ref:sample}. With these best fit parameters,
the normalized redshift probability distributions of the SDSS
galaxies and quasars sample and NVSS sample are shown together in
FIG.\ref{fig:reddis}.

Introducing the weight function $W_T(\chi)$ and $W_g(\chi)$ as
follows \cite{ref:ISW2}:
\begin{eqnarray}
&&W_T(\chi)=3a^2H(a)\frac{d\zeta(a)}{da},\\
&&W_g(\chi)=H(a)b(a)\frac{d N}{dz}D_+(a),
\end{eqnarray}
the line of sight integral for ISW temperature perturbation $\Delta
T/T_{CMB}$ and the density contrast of observed galaxies $\delta_g$
can be written as:
\begin{eqnarray}
&&\frac{\Delta T}{T_{CMB}}=\int_0^{\chi_H}d\chi
W_T(\chi)\frac{H_0^2\delta(k,a=1)}{k^2},\\
&&\delta_g=\int_0^{\chi_H}d\chi W_g(\chi)\delta(k,a=1),
\end{eqnarray}
which allow the expressions for the ISW-auto spectrum $C_{TT}(l)$,
the ISW-cross spectrum $C_{Tg}(l)$ and the observed galaxies-auto
spectrum $C_{gg}(l)$ to be written in a compact notation, applying
the Limber-projection \cite{ref:Limber} in the flat-sky
approximation, for simplicity:
\begin{eqnarray}
&&C_{TT}(l)=\int_0^{\chi_H}d\chi\frac{W_T(\chi)^2}{\chi^2}\frac{H_0^4P_{\delta\delta}(k=l/\chi)}{k^4},\\
&&C_{Tg}(l)=\int_0^{\chi_H}d\chi\frac{W_T(\chi)W_g(\chi)}{\chi^2}\frac{H_0^2P_{\delta\delta}(k=l/\chi)}{k^2},\\
&&C_{gg}(l)=\int_0^{\chi_H}d\chi\frac{W_g(\chi)^2}{\chi^2}P_{\delta\delta}(k=l/\chi),
\end{eqnarray}
where $P_{\delta\delta}(k)$ is the present matter power spectrum.
Here we take $P(k)\propto k^{n_s}T^2(k)$ and it can be normalized by
$\sigma_8=0.8$. $T(k)$ is the transfer function, and we adopt its
fitting form by \cite{ref:ISW2, ref:transferfunction}:
\begin{eqnarray}
&&T(q\equiv k/\Gamma
hMpc^{-1})=\frac{\ln[1+2.34q]}{2.34q}\left[1+3.89q+(16.2q)^2+
(5.47q)^3+(6.71q)^4\right]^{-0.25},
\end{eqnarray}
where the dimensionless quantity $\Gamma=\Omega_{m0}h$.

\begin{figure}[!htbp]
\includegraphics[width=12cm]{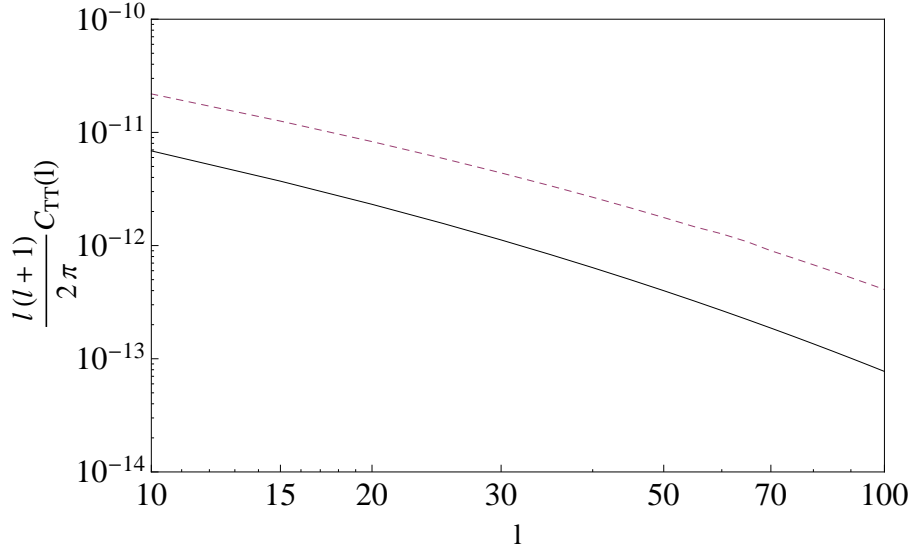}\\
  \caption{the ISW-auto power spectrum $C_{TT}$ vs. multipole order $l$ for $\Lambda \textmd{CDM}$
  model (solid line) and $\Lambda(t) \textmd{CDM}$ model (dashed line).
 }\label{fig:TT}
\end{figure}

\begin{figure}[!htbp]
\includegraphics[width=12cm]{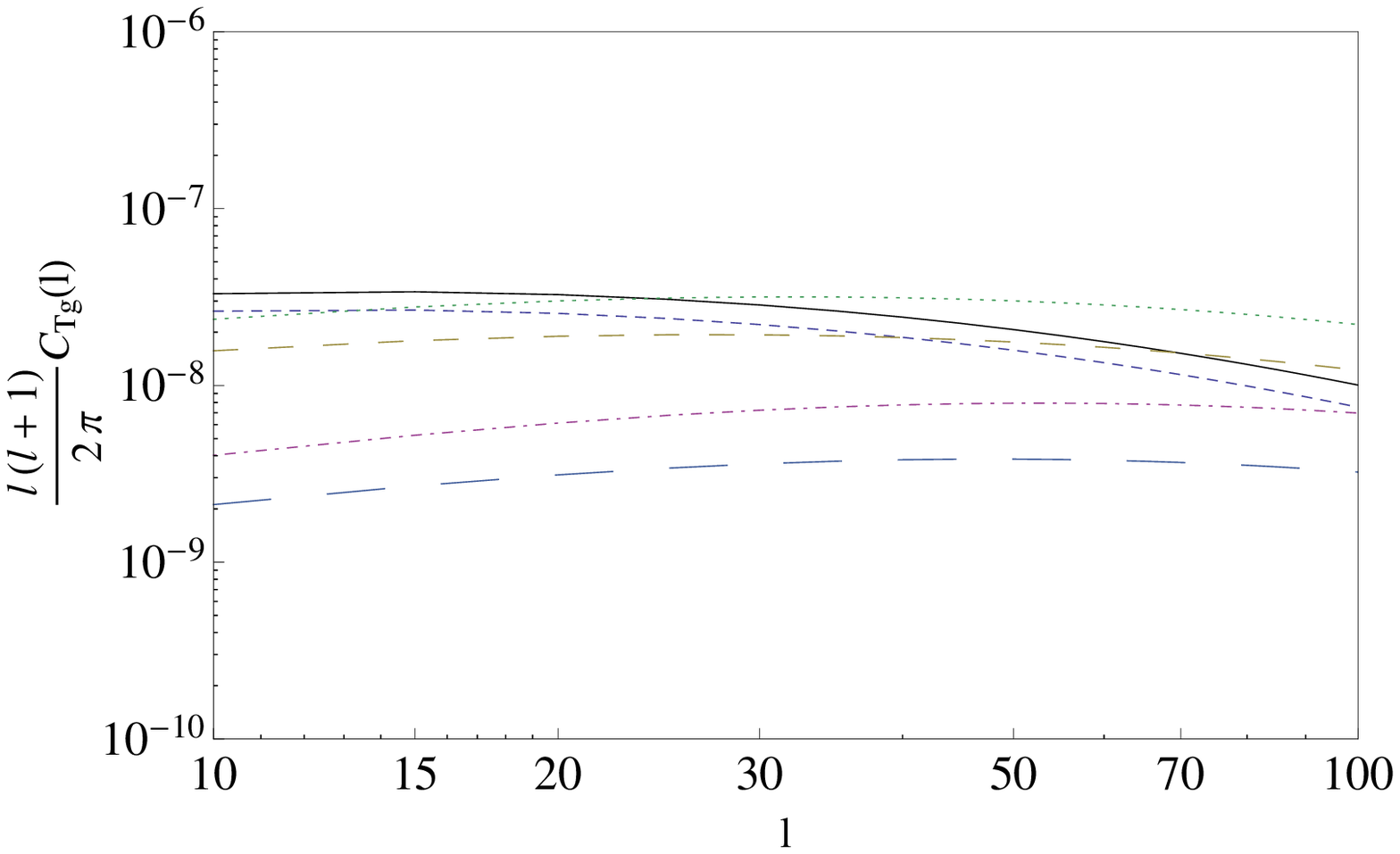}\\
  \caption{the ISW-cross power spectrum $C_{Tg}$ vs. multipole order $l$ for $\Lambda \textmd{CD}M$
  model: the short-dashed line is for the SDSS galaxies sample; the dashed line is for the NVSS sample; the long-dashed line is for
  the SDSS quasars sample; and $\Lambda(t) \textmd{CDM}$ model: the black line is for the SDSS galaxies sample; the dotted line is for
  the NVSS sample; the dot-dashed line is for the SDSS quasars sample.
 }\label{fig:Tg}
\end{figure}

\begin{figure}[!htbp]
\includegraphics[width=12cm]{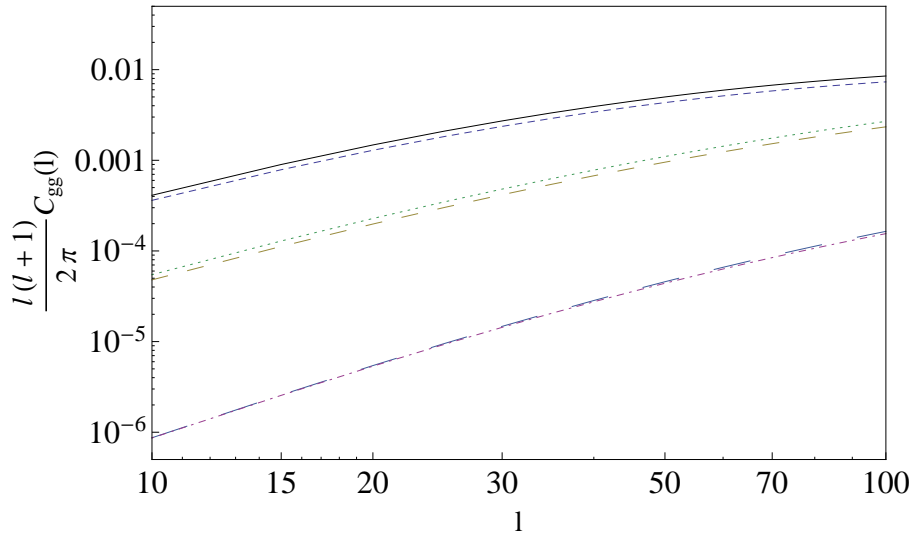}\\
  \caption{the observed galaxies-auto power spectrum $C_{gg}$ vs. multipole order $l$ for $\Lambda \textmd{CDM}$
  model: the short-dashed line is for the SDSS galaxies sample; the dashed line is for the NVSS sample; the long-dashed line is for
  the SDSS quasars sample; and $\Lambda(t) \textmd{CDM}$ model: the black line is for the SDSS galaxies sample; the dotted line is for
  the NVSS sample; the dot-dashed line is for the SDSS quasars
  sample.
 }\label{fig:gg}
\end{figure}

\begin{figure}[!htbp]
\includegraphics[width=12cm]{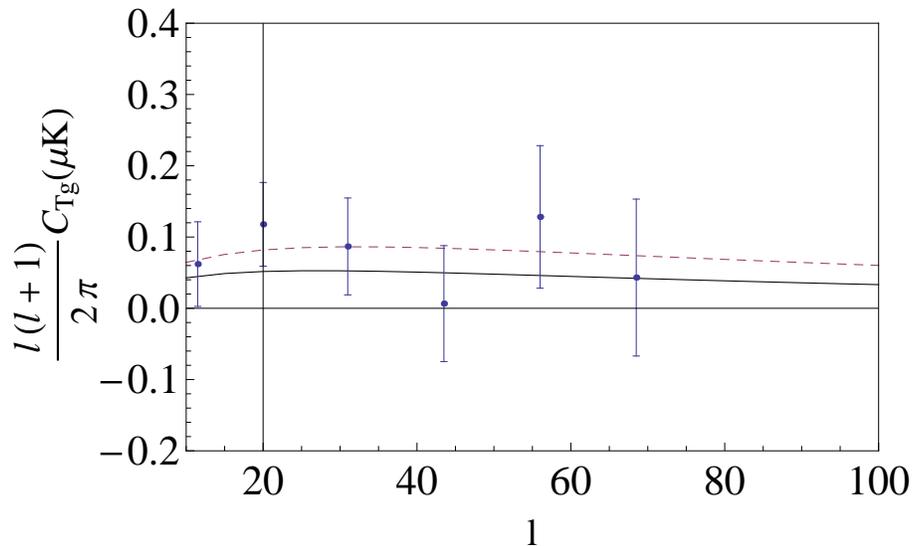}\\
  \caption{Comparison of the observational data of the ISW-cross power spectrum
  $C_{Tg}$ between ISW temperature and the radio galaxy catalogue from the
  NVSS \cite{ref:LSS3}
  and the theoretical evolutions of $C_{Tg}$: the dots with error bars denote the
  observed data with its corresponding uncertainty; the solid line is for the
  $\Lambda \textmd{CDM}$ model; the dashed line is for the $\Lambda(t)\textmd{CDM}$
  model. The well-measured temperature $T_{CMB}$ is 2.725K by \cite{ref:WMAP5,ref:T0}.
}\label{fig:NVSSTg}
\end{figure}

\begin{figure}[!htbp]
\includegraphics[width=12cm]{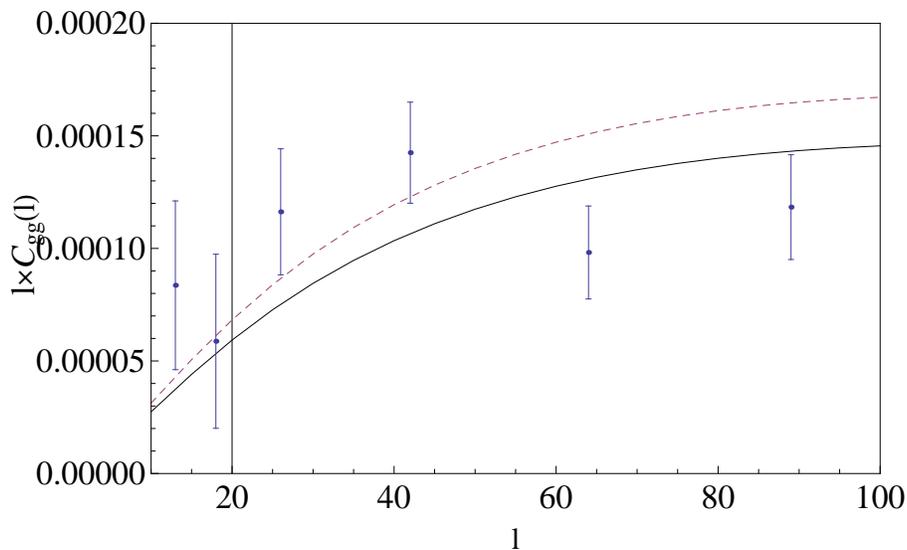}\\
  \caption{Comparison of the observational data of the observed galaxies-auto power spectrum
  $C_{gg}$ from the NVSS \cite{ref:LSS3}
  and the theoretical evolutions of $C_{gg}$: the dots with error bars denote the
  observed data with its corresponding uncertainty; the solid line is for the
  $\Lambda \textmd{CDM}$ model; the dashed line is for the $\Lambda(t)\textmd{CDM}$
  model.
 }\label{fig:NVSSgg}
\end{figure}

As shown in FIG. \ref{fig:TT}, \ref{fig:Tg}, \ref{fig:gg}, the power
spectrums, including the ISW-auto power spectrum $C_{TT}(l)$, the
ISW-cross power spectrum $C_{Tg}(l)$, and the observed galaxies-auto
power spectrum $C_{gg}(l)$, have been affected due to the presence
of the time varying vacuum energy. The amplitudes of the ISW-auto
power spectrums and ISW-cross power spectrums in
$\Lambda(t)\textmd{CDM}$ model are a few higher than those in the
cosmological constant model. There is a minor difference between the
observed galaxies-auto power spectrum amplitudes in the
$\Lambda(t)\textmd{CDM}$model and the $\Lambda \textmd{CDM}$ model.
The FIG.\ref{fig:Tg} and FIG. \ref{fig:gg} show respectively the
ISW-cross power spectrum $C_{Tg}(l)$ and the observed galaxies-auto
power spectrum $C_{gg}(l)$ are related to the observed samples with
different redshift ranges. In FIG. \ref{fig:Tg}, it's found that the
differences of the ISW-cross power spectrum $C_{Tg}(l)$ in the two
models become larger when the observed galaxies with wider redshift
ranges are used. Since the total CMB temperature-auto power spectrum
in $\Lambda \textmd{CDM}$ model is still consistent with the
observational results from WMAP, a few increase in the ISW-auto
power spectrum $C_{TT}(l)$ in $\Lambda(t) \textmd{CDM}$ model may
lead to the trouble between the total CMB temperature-auto power
spectrum in $\Lambda(t) \textmd{CDM}$ model and the observational
results from WMAP. As we have mentioned, it's difficult to directly
detect the amplitude of the ISW-auto power spectrum $C_{TT}(l)$
apart from the total CMB temperature anisotropies. However, it's
lucky that the ISW-cross power spectrum $C_{Tg}(l)$ provides an
effective measurement for the ISW effect. In FIG.\ref{fig:NVSSTg},
we show the observational data of the ISW-cross power spectrum
$C_{Tg}$ between ISW temperature and the radio galaxy catalogue from
the NVSS and its theoretical evolutions in $\Lambda \textmd{CDM}$
model and $\Lambda(t) \textmd{CDM}$ model. It's found that the
ISW-cross power spectrum $C_{Tg}$ in $\Lambda(t) \textmd{CDM}$ model
is still consistent with the observational results within $1\sigma$
errors. Although the amplitude of the ISW-auto power spectrum
$C_{TT}(l)$ in $\Lambda(t) \textmd{CDM}$ model is a few higher than
that in $\Lambda \textmd{CDM}$ model, its contribution to the total
CMB temperature power spectrum is still small and don't bring about
the disagreement between WMAP observations and the total CMB
temperature power spectrum. So, the WMAP measurements can not rule
the $\Lambda(t) \textmd{CDM}$ model out at present. In addition,
FIG.\ref{fig:NVSSgg} shows the comparison of the observational data
of the observed galaxies-auto power spectrum $C_{gg}$ from the NVSS
and the theoretical evolutions of $C_{gg}$ in the two models.

\section{Conclusion}

In summary, we have calculated the power spectrum of the ISW effect
in the time varying vacuum energy model. It's shown that the source
of the ISW effect is not only affected by the different evolutions
of the Hubble function $H(a)$ and the dimensionless matter density
$\Omega_m(a)$, but also by the different growth function $D_+(a)$,
all of which are changed due to the presence of matter production
term in the time varying vacuum model. In FIG. \ref{fig:growthrate}
and FIG. \ref{fig:D}, it's seen that the difference between the
growth function in $\Lambda(t)\textmd{CDM}$ model and that in
$\Lambda \textmd{CDM}$ model is not minor. What's more, the left
panel in FIG. \ref{fig:D} shows there is a crosspoint at $a=0.4$ for
the evolutions of the growth function in two models. The growth
function in $\Lambda \textmd{CDM}$ model is greater than that in
$\Lambda(t) \textmd{CDM}$ model before $a=0.4$. When $a>0.4$, the
growth function in $\Lambda \textmd{CDM}$ model is smaller than that
in $\Lambda(t) \textmd{CDM}$ model. However, since the ISW effect is
calculated by integrating time-dependent gravitational potential
from the time of last scattering to the present, the difference of
the ISW effect caused by the difference of the growth function in
the two models within the scale factor ranges $[0.4,1]$ can partly
counteract that caused by the discrepancy in their growth functions
at $a<0.4$ though the integration. Therefore, the difference of the
ISW effect in $\Lambda(t)\textmd{CDM}$ model and $\Lambda
\textmd{CDM}$ model is lessened to a certain extent due to the
integrated effect. As shown in FIG. \ref{fig:Tg}, there is not much
discrepancy between the availably observed quantities $C_{Tg}$ in
the two models for the SDSS galaxies and the sample from NVSS. For
the SDSS quasars with high redshift, the amplitudes of the ISW-cross
power spectrum in $\Lambda(t)\textmd{CDM}$ model and $\Lambda
\textmd{CDM}$ model obviously decrease. However, the relative
discrepancy between ISW-cross power spectrums in
$\Lambda(t)\textmd{CDM}$ model and $\Lambda \textmd{CDM}$ model for
the SDSS quasars is larger than those for the SDSS galaxies and the
sample from NVSS. Therefore, at present the ISW effect is weaker
than the growth history in distinguishing $\Lambda \textmd{CDM}$
model and $\Lambda(t)\textmd{CDM}$ model. However, it's worth
expecting the observational data of the ISW-cross power spectrum for
the observed galaxies with high redshift in the future.

\section*{Acknowledgments}
This work is supported by the National Natural Science Foundation of
China (Grant No 10703001), and Specialized Research Fund for the
Doctoral Program of Higher Education (Grant No 20070141034).

\end{document}